\documentclass[%
 reprint,
nofootinbib,
 amsmath,amssymb,
 aps,
]{revtex4-2}

\usepackage{graphicx}% Include figure files
\usepackage{dcolumn}% Align table columns on decimal point
\usepackage{bm}% bold math
\usepackage[colorlinks=true,citecolor=blue,urlcolor=blue]{hyperref}
\usepackage{multirow}
\usepackage{hypcap} %TJC added to use \capstart
\usepackage{float}

\newcommand{\Planck}{\textit{Planck}}

\newcommand{\lcdm}{$\Lambda$CDM}

\newcommand{\beq}{\begin{equation}}
\newcommand{\eeq}{\end{equation}}
\renewcommand{\arraystretch}{1.6}

\def\ee{\end{equation}}
\def\bea{\begin{eqnarray}}
\def\eea{\end{eqnarray}}
\def\bse{\begin{subequations}}

\usepackage{xcolor}
\usepackage{soul}%to allow comments to go on to next line

\begin{document}

%\preprint{APS/123-QED}

\title{A model-independent reconstruction of dark energy to very high redshift}% Force line breaks with \\

\author{Adam Moss}
 \email{adam.moss@nottingham.ac.uk}
\affiliation{%
 Centre for Astronomy \& Particle Cosmology, University of Nottingham, \\
 University Park, Nottingham, NG7 2RD, U.K.
}%

\author{Edmund J. Copeland}
 \email{edmund.copeland@nottingham.ac.uk}
\affiliation{%
 Centre for Astronomy \& Particle Cosmology, University of Nottingham, \\
 University Park, Nottingham, NG7 2RD, U.K.
}%

\author{Steven Bamford}
 \email{steven.bamford@nottingham.ac.uk}
\affiliation{%
 Centre for Astronomy \& Particle Cosmology, University of Nottingham, \\
 University Park, Nottingham, NG7 2RD, U.K.
}%

\author{Thomas J. Clarke}
 \email{ppxtjc.notts@gmail.com}
\affiliation{%
 Centre for Astronomy \& Particle Cosmology, University of Nottingham, \\
 University Park, Nottingham, NG7 2RD, U.K.
}%

\date{\today}% It is always \today, today,

\begin{abstract}
We provide a model-independent reconstruction of dark energy from $z=0$ to $ \gtrsim 10^5$. We parameterise the model by a perfect fluid with a series of physically well-motivated bins in energy-density, such that the equation of state is always $-1 \le w \le 1$. Our method is capable of describing a range of theoretical models with smooth modifications to the expansion history. Combining the latest CMB, BAO, SN and local $H_0$ measurements, we obtain a large improvement of $\Delta \chi^2=41.3$ over \lcdm, at the expense of 33 additional parameters in the fit, with dark energy contributing significantly between $z \sim 10^4 - 10^5$, and intriguingly with a sound speed $c_s^2 \sim 1/3$. A significant part of the $\Delta \chi^2$ improvement comes from \Planck\ + Atacama Cosmology Telescope (\textsc{Act}) data, alleviating tension between them within \lcdm. We apply a correlation prior to penalise models with unnecessary degrees of freedom, and find no  preference  for deviations  from \lcdm\  at  late-times, but moderate Bayesian evidence of an early dark energy (EDE) component. Although the model has a large amount of freedom, it is unable to reduce  $S_8 \equiv \sigma_8 (\Omega_\mathrm{c} / 0.3)^{0.5}$ below that of \lcdm, to bring about full concordance with large-scale structure data.
\end{abstract}

\maketitle

{\em Introduction.---}%
The discrepancy between high and low redshift measurements of the Hubble constant, $H_0$, is one of the main challenges of the standard \lcdm\ model. The \Planck\ CMB value, which is most sensitive to physics at $z \gtrsim 1000$, finds $H_0=67.27\pm0.60$ km/s/Mpc~\citep{Aghanim:2018eyx}, while the latest distance-ladder measurement from the Supernovae H0 for the Equation
of State (SH0ES) project, finds $H_0=73.2\pm1.2$ km/s/Mpc~\citep{Riess:2020fzl}\footnote{In this paper we use the~\citet{Riess:2019cxk} SH0ES value of $H_0=74.03\pm 1.42$ km/s/Mpc. We do not expect this to significantly change our conclusions.}. Other local measurements, such as strong-lensing time delays~\citep{Wong:2019kwg} and megamasers~\citep{Pesce:2020xfe} support high values of $H_0$, although different analyses give somewhat smaller values (e.g.~\citep{Freedman:2019jwv, Birrer:2020tax}). Overall, the level of tension between \Planck\ CMB and local $H_0$ measurements can be as high as $4-6\sigma$, depending on the exact data used.

There are a plethora of models that attempt to solve this problem (see~\citep{DiValentino:2021izs} for a comprehensive review). All must satisfy certain geometric constraints~\citep{Knox:2019rjx}, such as the CMB acoustic scale, $\theta_{\star} \equiv r_{\star} / D (z_{\star})$, which is accurate to $0.05\%$. Broadly, models can be classified into early and late Universe solutions, with the latter modifying $H(z)$ at low redshift, in order to keep the angular diameter distance to recombination, $D (z_{\star})$, fixed. The former reduces the sound horizon, $r_{\star}$, at recombination, to accommodate a smaller $D (z_{\star})$ that results from a higher $H_0$. 

Modifications to dark energy have been used in both contexts. Proposals include the presence of an Early Dark Energy (EDE) component (e.g.~\citep{Poulin:2018cxd, Agrawal:2019lmo, Smith:2019ihp, Niedermann:2019olb}), comprising $\sim 10\%$ of the energy-density at $z \sim 5000$. At late times, it has been shown that dynamical dark energy can alleviate the Hubble tension (e.g~\citep{Zhao:2017cud, Wang:2018fng}). However, late-time solutions have difficulty reconciling Baryon Acoustic Oscillation (BAO) data. Here we attempt to be as agnostic as possible, the only assumption being that dark energy has a conserved, perfect fluid description, whose equation of state satisfies $-1 \le w \le 1$. 

There have been many studies reconstructing dark energy, but most of these apply to the late Universe and attempt to reconstruct $w(z)$. In this letter  we perform a model-independent reconstruction from $z=0$ to $\gtrsim 10^5$. Rather than using $w(z)$, we fit for the dark energy fraction, $f_\text{DE}(z) \equiv \rho_\text{DE}(z)/\rho_\text{crit}(z)$. Although they are linked through the conservation equation, obtaining $f_\text{DE}(z)$ requires integrating $w(z)$. If there was a preference for a non-zero $f_\text{DE}(z)$ at high redshift, this would require severe fine-tuning of $w(z)$. Our model is similar to Acoustic Dark Energy (ADE~\citep{Lin:2019qug} and the reconstruction method of~\citep{Hojjati:2013oya}, but we use a much finer set of bins, enabling us to reconstruct a wider range of cosmologies.

There are further, weaker cosmological tensions. The value of $S_8 \equiv \sigma_8 (\Omega_\mathrm{m} / 0.3)^{0.5}$, where $\sigma_8$ is the matter clustering amplitude on scales of $8h^{-1}\,{\rm Mpc}$, and $\Omega_\mathrm{m}$ is the matter density, is in $2-3\sigma$ tension between \Planck\ and cosmic-shear results from DES~\cite{DES:2021wwk} and KiDS~\cite{Heymans:2020gsg}. As explained in~\citep{Ivanov:2020ril, DAmico:2020ods, Jedamzik:2020zmd}, most EDE models have a difficult time resolving both the $H_0$ and $S_8$ tensions. %An issue is that a higher $\Omega_\mathrm{m} h^2$ is required to compensate effects in the CMB, and this leads to a higher $S_8$. 
In addition, there is further tension between \Planck\ and  Atacama Cosmology Telescope (\textsc{Act}) data. As discussed in~\citet{Aiola:2020azj}, \textsc{Act} data alone prefers a lower first TT peak and a lower (higher) TE (EE) spectrum, at the $2.5\sigma$ level. In a joint fit, cosmological parameters are shifted from their preferred values for each dataset, most noticeably in $n_\mathrm{s}$ and $\Omega_\mathrm{b} h^2$. We address whether these additional tensions can also be resolved with much more freedom in $f_\text{DE}(z)$.

{\em Axion EDE Model.---}%
We firstly briefly review the axion EDE model (e.g.~\cite{Poulin:2018dzj}), as it shares several similarities with our approach, and provides a useful baseline. The axion potential is given by $V(\phi)\propto \left(1-\cos{[\phi / f]}\right)^n$, where $\phi$ is the field value, $f$ is the decay constant and $n$ is a (not-necessarily integer) constant. The choice $n=1$ corresponds to the standard axion potential. At early times the field acts like a cosmological constant, after which it  oscillates around the potential minima with an effective equation of state $w_n=(n-1) / (n+1)$.

The axion model can be approximated by a fluid governed by 4 parameters, $\left\{ a_\mathrm{c}, \Omega_{a}(a_c), w_n, \theta_i \right\}$. The first, $a_\mathrm{c}$ is the critical value of the scale factor at which the fluid transitions away from a cosmological constant, and $\Omega_a({a_c})$ is the fractional energy density at this time. In the fluid approximation  the energy density evolves as~\cite{Poulin:2018dzj}
\begin{equation}
 \Omega_a(a)= \frac{2 \Omega_{a}(a_c)}{\left(a/a_c\right)^{3( w_n+1)}+1},\label{eq:omegaFit}
 \end{equation}
 with an equation of state 
 \begin{equation}
     w_a(a) = \frac{1+w_n}{1+(a_c/a)^{3(1+w_n)}}-1\,.
     \label{eq:wphi}
 \end{equation}
 Finally, $\theta_i$ is the initial field value and determines the time and scale dependence of the effective sound-speed, $c_\mathrm{s}$~\cite{Poulin:2018dzj}.
 % This is given by 
%\begin{equation}
%    c_\mathrm{s}^2 = \frac{2a^{(2-6w_n)}\bar{\omega}_0^2(n-1)^2+k^2}{2a^{(2-6w_n)}\bar{\omega}_0^2(n+1)^2+k^2} \,.
%\end{equation}
%where $\bar{\omega}_0$ is the oscillation frequency of the field. 
Note that, for $n\to\infty$, $w_n\to1$ and $c_\mathrm{s}^2\to1$. For the best-fit axion model $n \approx 3$, with $c_\mathrm{s}^2 \approx 0.7$ over the relevant times and scales of interest~\cite{Smith:2019ihp}.

Once the perturbation equations are specified, the full evolution of the fluid can be calculated. In the synchronous gauge the equations for the density contrast, $\delta_\text{a}$ and heat-flux $u_\text{a}$, for the mode $k$, are~\cite{Weller:2003hw},
\begin{eqnarray} \label{eqn:perts}
    \dot{\delta_\text{a}}&=&-\bigg[k u_\text{a}+(1+w_\text{a})\frac{\dot{h}}{2}\bigg]-3 {\cal H} (c_\mathrm{s}^2-w_\text{a})\left( \delta_\text{a} + 3 {\cal H} \frac{u_\text{a}}{k} \right) \nonumber\\
     & & - 3 {\cal H} \frac{ \dot{w}_\text{a}}{(1+w_\text{a}) } \frac{ u_\text{a}}{k}  \,,\\ 
    \dot{u_\text{a}}&=&-(1-3c_\mathrm{s}^2){\cal H}u_\text{a}+  \frac{\dot{w}_\text{a}}{(1+w_\text{a}) } u_\text{a}  + k c_\mathrm{s}^2\, \delta_\text{a}\,,
\end{eqnarray}
where ${\cal H} = a H$, $c_\mathrm{s}$ is defined in the rest-frame of the fluid, and the heat-flux, $u_\text{a} \equiv (1+w_\text{a}) v_\text{a}$, is favoured over the velocity, $v_\text{a}$, for numerical stability when $w_\mathrm{a}\approx-1$. 

{\em Model-independent approach.---}% 
There are many theoretical models that modify the expansion history at early times, so it is desirable to develop a model-independent approach. To do this, we modify the Friedmann equation with a set of $N$ non-interacting fluids, each with energy density $\Omega_i$,
\begin{equation} \label{eq:H0:friedmann}
    H^2 (a) = H_0^2 \left[\Omega_\mathrm{\Lambda CDM} (a) +  \sum_{i=1}^N  \Omega_i(a) \right]\,,
\end{equation}
where $\Omega_\mathrm{\Lambda CDM}(a)$ is the total \lcdm\ density, consisting of matter, radiation and a cosmological constant. For each additional fluid, we choose a functional form for the equation of state such that it scales like a cosmological constant before a transition scale, $a_i$, and as a stiff fluid after,
\begin{equation}
    w_i(a) = \frac{2}{1+ (a_i/a)^{\beta}}-1,
\end{equation}
where $\beta >0$ is a parameter that sets the speed of the transition. The energy density of each component is then
\begin{equation} \label{eqn:omega}
 \Omega_i(a)=  \Omega_{i} \left( \frac{2a_i^\beta}{a^\beta + a_i^\beta} \right)^{6/\beta} \,,
 \end{equation}
where $ \Omega_{i}$ is the density at the transition scale. The case $\beta=6$ corresponds to an axion fluid with $w_n=1$.

We call this the {\em spike model}, since each component has a maximum energy-density, relative to the background, at $a_i$. These can be thought of as a well-defined set of basis modifications to $H^2 (a)$, since they obey $-1 \le w_i \le 1$ by construction.  In our analysis, we choose a fixed set of $a_i$, logarithmically spaced from $a=5 \times 10^{-6}$ to $1$. This means our reconstruction applies to both early and late-time dark-energy. The lower limit is chosen as there is little sensitivity in CMB data to EDE for $a \lesssim 5 \times 10^{-6}$. 

Perturbations are modelled by treating the $N$ fluids as a single effective fluid with equation of state, $w_\mathrm{eff} =  \sum_i \Omega_i w_i /  \sum_i \Omega_i$, which is similarly bounded by $-1 \le w_\mathrm{eff} \le 1$. We use the same perturbation equations as the axion fluid, but assume the rest-frame sound-speed is constant, with $0 \le c_{s}^2 \le 1$. Although our model is similar to ADE~\cite{Lin:2019qug}, they only consider a single component with a variable transition scale. 
 \begin{figure}
% Produced with best_fit_axion.ipynb
\capstart
\includegraphics[width=\columnwidth]{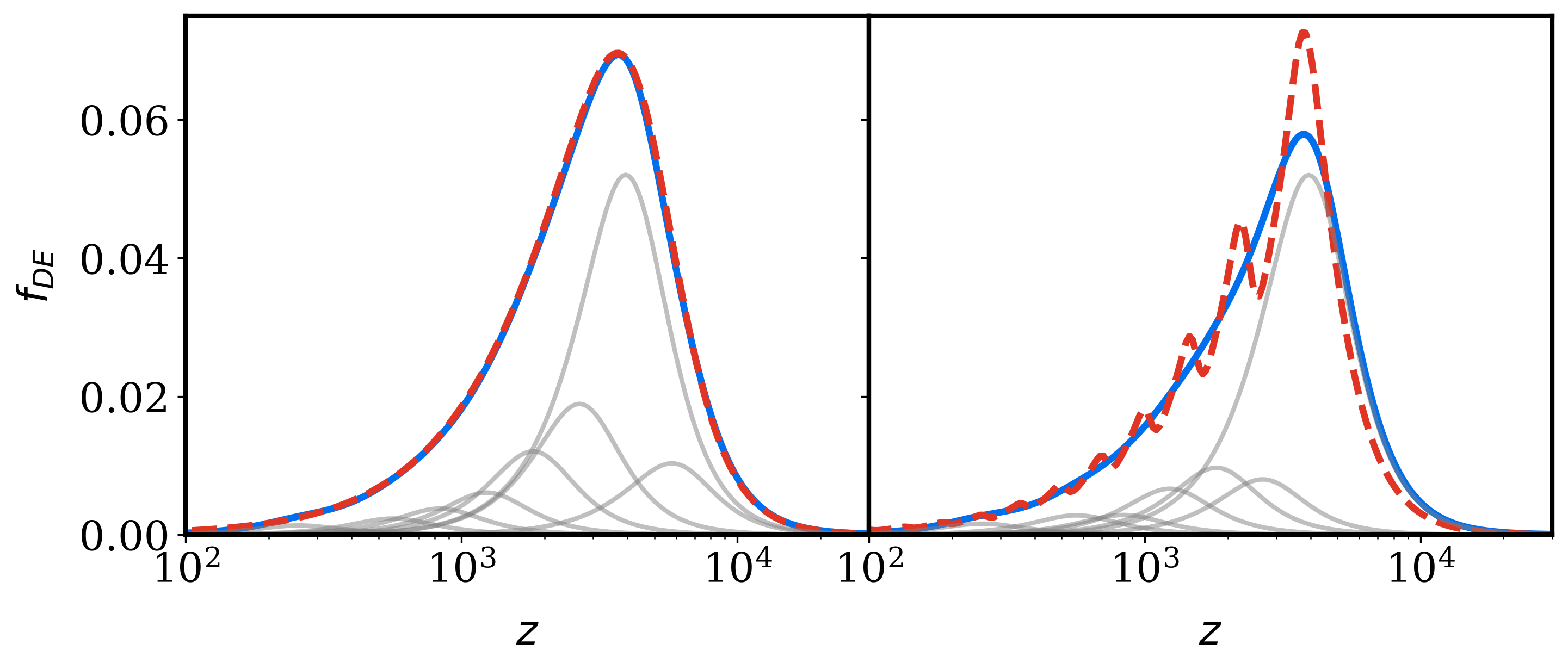}
\caption{Reconstruction of the best-fit axion fluid (left) and full scalar-field evolution (right). The axion $f_\text{DE}(z)$ is shown by the dashed red curve, the reconstruction by the solid blue curve, and each of the fitted spike components in solid grey. }
\label{fig:axion_reconstruction}
\end{figure}

 We find that $N=32$ components with $\beta=6$ is sufficient to reconstruct a range of theoretical models with `smooth' modifications to the expansion history, such as the axion fluid and tracking models of quintessence, with minimal bias. As an example, the left-hand panel of Fig.~\ref{fig:axion_reconstruction} shows the dark energy fraction, $f_\text{DE}(z)$, of the best-fit axion fluid to the baseline+ext data combination, defined in the following section. The amplitudes of the spike model, $\Omega_{i}$, are then fitted to minimise the least-squares fit to $f_\text{DE}(z)$, with  $\sim 7$ non-zero components required. The sound speed is chosen to minimise the least-squares fit to $C^{TT}_{\ell}$ up to $\ell=3000$, with the optimal value found to be $c_s^2=0.68$.  This `axion mimic' model has $\chi_\mathrm{axion}^2 - \chi_\mathrm{mimic}^2 = -3.3$ when evaluated with the full likelihood code, which can be attributed almost entirely to the variable sound speed in the axion model.
 
 An example of where the reconstruction fails is the full scalar-field evolution of the axion. This is shown in the right-hand panel of  Fig.~\ref{fig:axion_reconstruction}. Although the fitted components match the overall behaviour of $f_\text{DE}(z)$, they are unable to recover the oscillatory behaviour. This would require an even larger number of components and higher $\beta$, which would make a reconstruction using cosmological data very challenging.

{\em Data and Results.---}% 
We perform a Markov Chain Monte Carlo (MCMC) analysis of the \lcdm\,, axion fluid and spike models using the public \textsc{Cobaya}~\citep{Torrado:2020dgo} and \textsc{Camb} codes~\citep{Lewis:1999bs}. We find some of the posterior distributions are lightly multi-modal and chains have long mixing-times, so incorporate the ensemble sampler \textsc{emcee}~\citep{emcee} to sample over the model parameters ${\mathbf P}$, which can improve autocorrelation times over traditional MCMC methods. We run 100 walkers in the ensemble, using a combination of the affine invariant stretch~\cite{2010CAMCS...5...65G} and differential evolution moves. The minimum $\chi^2$ is then found by BOBYQA minimisation, using the chain best-fit as an initial guess~\cite{BOBYQA}. Performing this step is especially important with a large number of parameters, as the best-fit from the chain can be significantly worst due to sampling error. We use the following datasets:

{\bf Baseline:} We use \Planck\ 2018 data~\cite{Aghanim:2018eyx} in combination with BAO data from BOSS DR12~\cite{Alam:2016hwk}, 6dFGS~\cite{beutler:2011:6dfgsbao} and SDSS-MGS~\cite{ross:2014:sdssbao}. The \Planck\ likelihoods used are the TT, TE and EE spectra at $\ell\ge 30$, the low-$\ell$ likelihood using the \textsc{Commander} component separation algorithm~\cite{Akrami:2018mcd}, the low$-\ell$ EE likelihood from the \textsc{SimAll} algorithm, and lensing~\citep{Aghanim:2018oex}. In order to reduce the number of MCMC parameters, we use the foreground marginalized `lite' version of the \Planck\ likelihood.

{\bf Ext:} We include the SH0ES $H_0$ measurement from ~\citep{Riess:2019cxk} and high $\ell$ CMB data from \textsc{Act} DR4~\citep{Aiola:2020azj}. For \textsc{Act} we exclude the large scale temperature data, to minimise double counting when combining with \Planck\footnote{It is shown in~\cite{Hill:2021yec} that increased accuracy settings are required in \textsc{Camb} in order to give full convergence in the \textsc{Act} $\chi^2$. Our analysis is performed at the default settings, as these higher accuracy settings require an order of magnitude longer runtime. We have checked that for a sample of models, the absolute \textsc{Act} $\chi^2$ values are accurate to $2-3$, and the relative $\chi^2$ differences with respect to \lcdm\ are even smaller.}. In addition, we include the Pantheon SN sample~\cite{Scolnic:2017caz}.

{\bf $S_8$ prior:}  EDE models tend to have an increased value of $S_8 $ compared to \lcdm. To quantify this, we perform additional runs with the inclusion of a $S_8$ prior, using the DES value of $S_8 = 0.776 \pm 0.017$. We use a prior rather than the full likelihood again to reduce the number of MCMC parameters, and this has been shown to be a good approximation for the axion~\cite{Hill:2020osr}.

We first produce runs for $\Lambda$CDM and the axion fluid, assuming flat priors on the base $\Lambda$CDM parameters, $\left\{ H_0, \Omega_\mathrm{c} h^2, \Omega_\mathrm{b} h^2, n_\mathrm{s}, \log(10^{10} A_\mathrm{s}), \tau  \right\}$, and $\left\{ z_\mathrm{c}, f_\mathrm{DE}(z_c), w_n, \theta_i \right\}$ for the axion. We use $H_0$ rather than $\theta_{\star}$ as a base parameter, to ensure any preference for larger $H_0$ is not prior driven (\cite{Millea:2018bko} demonstrates this issue when reconstructing the ionization fraction). %There are an additional two nuisance parameters explicitly fitted for, the Planck absolute calibration and \textsc{Act} polarization efficiency. 
As per the \Planck\ analysis, neutrinos are modelled as 2 massless species, and one massive species with $m_{\nu}=0.06\,{\rm eV}$. 
\begin{figure}
\capstart
\includegraphics[width=\columnwidth]{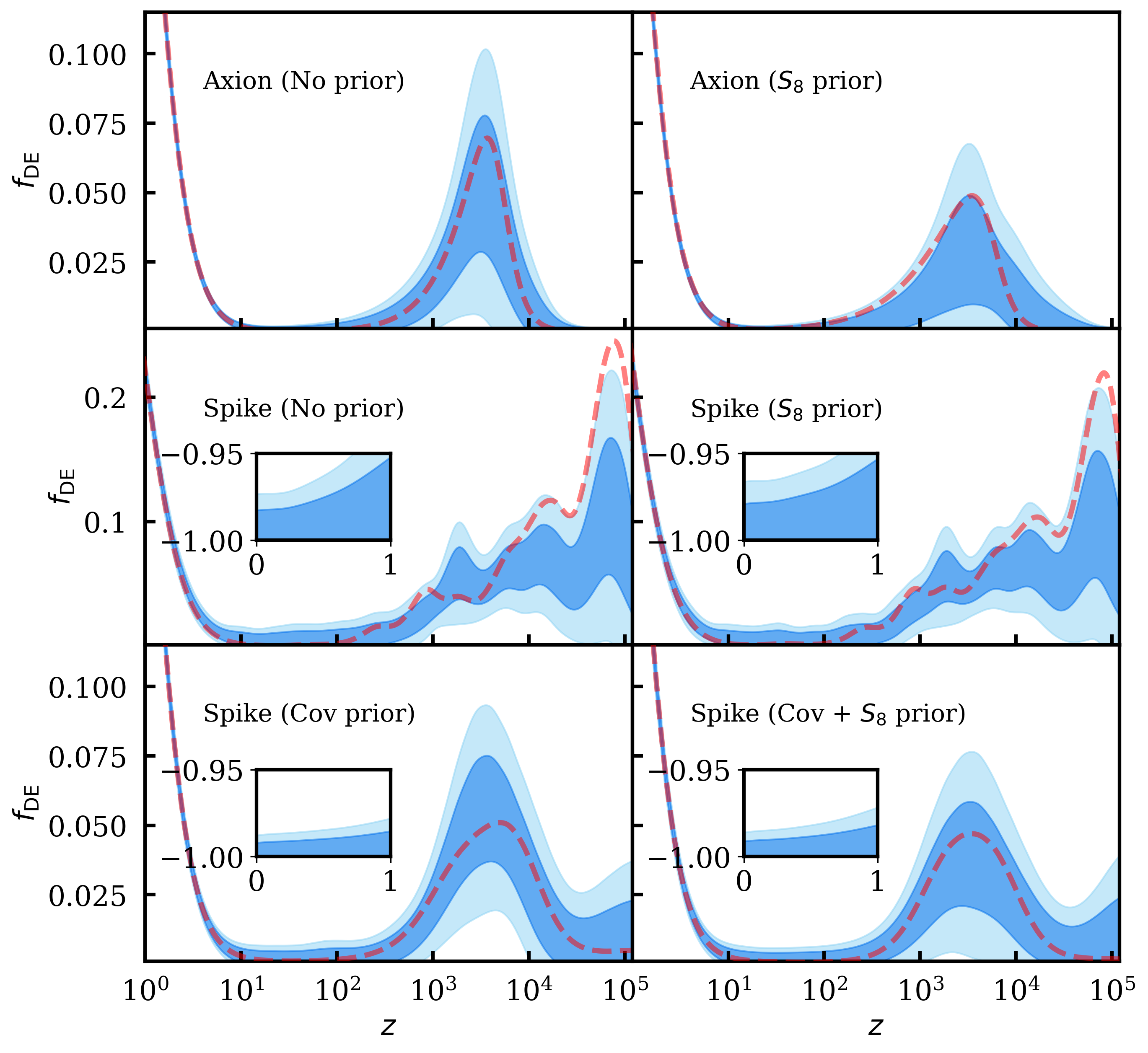}
\caption{$f_\text{DE}(z)$ for the axion fluid (top-row) and the spike model without (middle) and with (bottom) a covariance prior.  On the left (right) we show results without (with) a DES $S_8$ prior, otherwise using the baseline+ext dataset. 1 and 2$\sigma$ confidences are indicated by the dark and light blue regions, and the best-fit by the dashed line. The dark energy includes the cosmological constant contribution. The inset shows the resulting late-time $w(z)$ reconstruction.}
\label{fig:fede}
\end{figure}

\def\arraystretch{1.2} %vertical space
\begin{table*}
\capstart
\centering
\resizebox{\textwidth}{!}{\begin{tabular} {| l || c | c | c | c | } \hline
 Parameter & \lcdm &  Axion Fluid & Spike & Spike (+ Covariance Prior) \\ \hline\hline
$H_0$ & $68.48\pm 0.32$ (68.44) & $70.03^{+0.81}_{-1.1}$ (70.95) & $72.25^{+0.93}_{-1.2}$ (73.59) & $70.9^{+1.0}_{-1.3}$ (71.29)\\
$\Omega_\mathrm{m}$ & $0.3001\pm 0.0041$ (0.3006) & $0.2975^{+0.0044}_{-0.0049}$ (0.2950) & $0.3027^{+0.0062}_{-0.0055}$ (0.2978) & $0.2948\pm 0.0054$ (0.2952)\\
$n_\mathrm{s}$ & $0.9729\pm 0.0030$ (0.9728) & $0.9810^{+0.0060}_{-0.0073}$ (0.9834) & $0.9703\pm 0.0083$ (0.9636) & $0.9805^{+0.0081}_{-0.0063}$ (0.9833)\\
$c_\mathrm{s}^2$ & - & - & $0.334^{+0.021}_{-0.039}$ (0.3125) & $0.401^{+0.10}_{-0.090}$ (0.4153)\\
$w_\mathrm{n}$ & - & $0.475^{+0.087}_{-0.18}$ (0.3523) & - & -\\
$z_\mathrm{c}$ & - & $10240^{+2000}_{-8000}$ (5460) & - & -\\
$f_\mathrm{EDE}(z_\mathrm{c}) $ & - & $0.0272^{+0.0097}_{-0.021}$ (0.03609) & - & -\\
$S_8$ & $0.8075\pm 0.0077$ (0.8073) & $0.814\pm 0.010$ (0.8133) & $0.8182\pm 0.0099$ (0.8183) & $0.812^{+0.011}_{-0.0094}$ (0.8151)\\ \hline
$\chi^2_\mathrm{H0}$ & 15.5 & 4.7 ({\bf -10.8}) & 0.1 ({\bf -15.4}) & 3.7 ({\bf -11.8})\\
$\chi^2_\mathrm{Planck}$ & 1017.0 & 1020.0 ( 3.0) & 1009.2 ({\bf -7.8}) & 1018.3 ( 1.3)\\
$\chi^2_\mathrm{ACT}$ & 240.7 & 235.3 ({\bf -5.4}) & 225.3 ({\bf -15.4}) & 234.4 ({\bf -6.3})\\
$\chi^2_\mathrm{S8}$ & 3.4 & 4.8 ( 1.4) & 6.2 ( 2.8) & 5.3 ( 1.9)\\ \hline
$\chi^2_\mathrm{data}$ & 2316.7 & 2305.9 ({\bf -10.8}) & 2281.4 ({\bf -35.4}) & 2302.8 ({\bf -14.0})\\
$\chi_\mathrm{prior}^2$ & 0.0 & 0.0 & 0.0 & 3.8\\
$\Delta \ln E $  & - & - & -  & 5.0  \\ \hline
\end{tabular}}
\caption[Mean and best-fit parameter values for the \lcdm, axion fluid and spike model]{Mean and best-fit parameter values for the \lcdm, axion fluid and spike models, for the baseline+ext+$S_8$ dataset. Consistent parameters and $\chi^2$ values have been suppressed.}
\label{table:parameters}
\end{table*}

For the axion fluid, we find a $\Delta \chi^2=-16.2$ improvement over  \lcdm\ for the baseline+ext dataset, which is consistent with other studies\footnote{A slightly improved fit is possible when considering the full scalar-field evolution~\citep{Smith:2019ihp}.}. The bulk of this ($\Delta \chi_\mathrm{H0}^2=-12.3$) comes from the SH0ES measurement, with a smaller contribution of $\Delta \chi_\mathrm{ACT}^2=-5.2$. In the recent analysis of~\cite{Hill:2021yec}, they find a preference for a scalar-field axion when combining large-angle  \Planck\ ($\ell < 650$) and \textsc{Act} data ($\Delta \chi^2_\mathrm{ACT}=-16.1$), driven primarily by an improved fit to the \textsc{Act} EE spectrum. However, this effect largely disappears when combined with the full \Planck\ data ($\Delta \chi^2_\mathrm{ACT}=-6.1$), similar to our findings. The axion model is therefore unable to fully account for the \Planck\ + \textsc{Act} tension (see also~\cite{Poulin:2021bjr}).

The resulting $f_\text{DE}(z)$ is shown in the top-left panel of Fig.~\ref{fig:fede}, with a marginalised value of $f_\mathrm{DE}(z_\mathrm{c}) = 0.050^{+0.023}_{-0.033}$, occurring at $z_\mathrm{c} = 5417^{+470}_{-2000}$. Analysing the posterior distributions, one can observe a bi-modality in $w_n$, with peaks at $w_n \approx 1/2$ and $w_n \approx 1/3$. The former has a higher likelihood, and the latter is correlated with a higher $z_\mathrm{c}$ and lower $f_\mathrm{DE}(z_\mathrm{c})$, which will be relevant when we come to interpret the results from the spike model.

The result of applying the DES prior is shown in the top-right panel of Fig.~\ref{fig:fede} and Table.~\ref{table:parameters}. The $\Delta \chi^2$ improvement over  \lcdm\ is now reduced to $-10.8$, primarily due to a poorer fit to \Planck\ ($\Delta \chi^2_\mathrm{Planck}=+3.0$) and a slightly higher $S_8$ value. As noted in~\citep{Murgia:2020ryi}, however, although the axion does not bring about concordance, it does not significantly worsen the fit compared to \lcdm.

For the spike model we assume flat priors on $ 0 \leq c_\mathrm{s}^2 \leq 1$ and $ -5 \leq \Delta_i  \leq 0$, where $\Delta_i = \log_{10} \left[ \Omega_{i} / \Omega_\mathrm{\Lambda CDM} (a_i) \right]$. We use a log transform due to the large dynamical range -- near $z \sim 5000$ the data requires $\Delta_i \lesssim -4$ to be indistinguishable from \lcdm, but the upper $2\sigma$ limits can be as high as $\Delta_i \approx -1$. As discussed in the next section, an unbounded prior is also required when applying a Gaussian correlation prior. 

One issue with this parameterisation is that the likelihood is slowly varying for low $\Delta_i$, which means the posterior is dominated by large, flat regions with a relatively good likelihood. The best-fitting models occupy a much smaller volume, and although they have an improved $\chi^2$, the chains mix slowly  with a long auto-correlation time. In our runs we have sampled for $15000$ iterations but still observe some slowly varying features in the trace plots. We have checked our results aren't significantly affected by performing different sample splits along the chain, and observe similar empirical means and variances.    

With increased freedom in fitting the expansion history, we find a large $\Delta \chi^2=-41.3$ improvement over  \lcdm\ for the baseline+ext dataset, and $\Delta \chi^2=-35.4$ with the inclusion of the DES prior. These represent improvements of $\Delta \chi^2=-25.1$ and $\Delta \chi^2=-24.6$ over the axion fluid. The best-fit $H_0$, shown in Table.~\ref{table:parameters}, is now almost entirely consistent with the SH0ES value. What is perhaps more intriguing is a substantial improvement to the joint \Planck\ + \textsc{Act} data, with $\Delta \chi^2_\mathrm{Planck}=-7.8$ and $\Delta \chi^2_\mathrm{ACT} = -15.4$. 

\begin{figure*}
\capstart
\includegraphics[width=\textwidth]{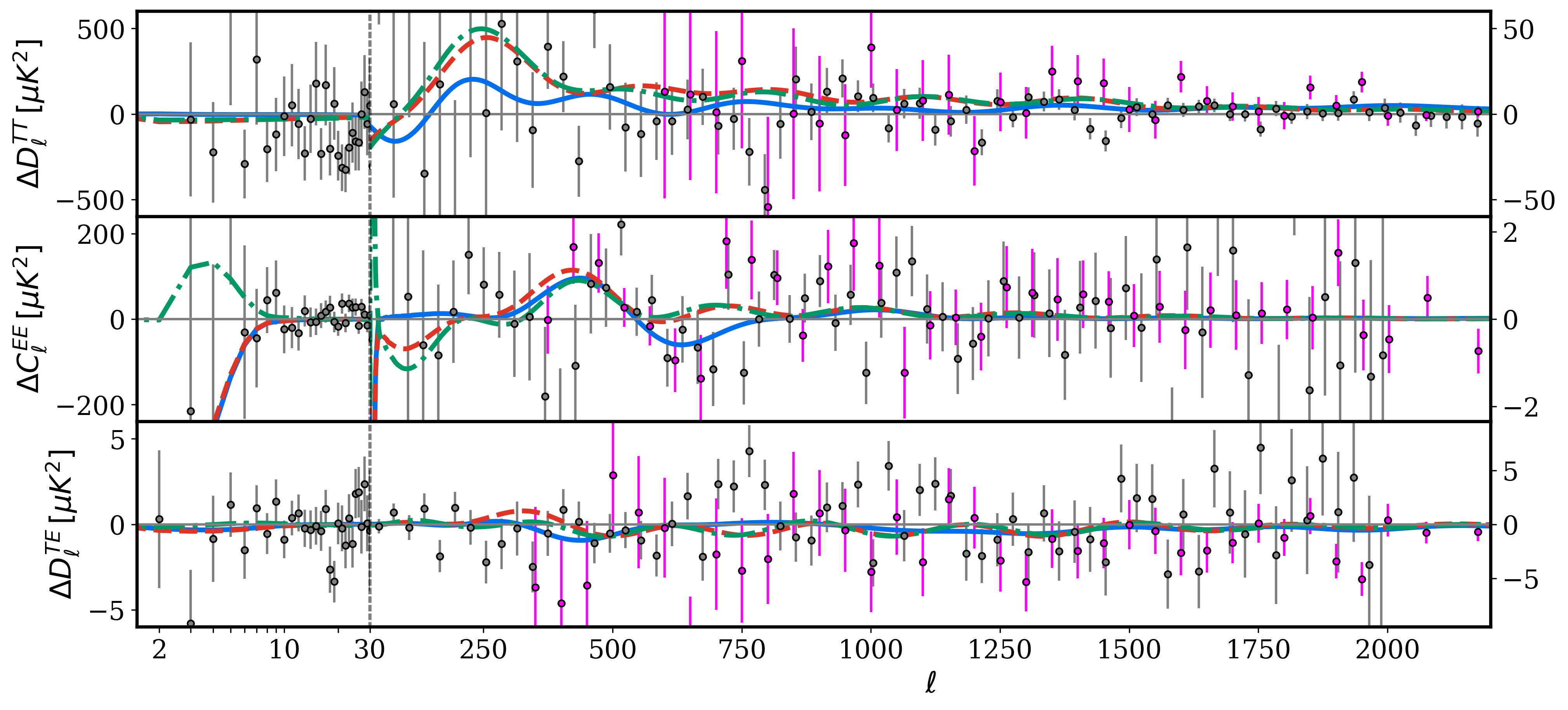}
\caption{Residual TT (top), EE (middle) and TE (bottom) CMB power spectra for the best-fit spike (solid blue), spike with covariance prior (dashed red), and axion fluid (dot-dash green), derived using baseline+ext+$S_8$ data, relative to \lcdm\ using  only baseline+\textsc{Act} data. The \Planck\ 2018 data~\cite{Aghanim:2018eyx} is shown in grey and the \textsc{Act} DR4~\citep{Aiola:2020azj} data in magenta.}
\label{fig:cls}
\end{figure*}

In Fig.~\ref{fig:cls} we show residuals of the TT, TE and EE power spectra for the axion and spike models, derived using baseline+external+$S_8$ data, versus \lcdm\ using  {\em only} baseline+\textsc{Act} data. In contrast to the axion, the spike model is able to  fit the dip in the EE spectrum at $\ell \sim 650$. It is also able to better fit the residuals in TT data in the range $ 500 < \ell < 1200$. %which result from parameter shifts in \lcdm. 

The reconstruction of $f_\text{DE}(z)$ for the spike model with minimal prior assumptions is shown in the middle panel Fig.~\ref{fig:fede}, along with an inset of $w(z)$ at late-times. The reconstruction shows no preference for deviations from \lcdm\ at late-times, but a strong preference for EDE in the range $z  \sim 10^{3} - 10^{5}$. Interestingly, $f_\text{DE}(z)$ is significantly different to the axion model, with a large peak at $z \sim 10^{5}$, after which it decays, on average, only slightly faster than the background. There is also a strong preference for a lower sound speed of $c_\mathrm{s}^2 \sim 1/3$. This means that the EDE is behaving like, but not exactly the same as, additional radiation. Despite a large $f_\text{DE}(z)$ at high $z$, it has a diminishing effect on the CMB. The visibility averaged $r_{\star}$, which weights the relative contribution by redshift, is  peaked near $z_{\star}$~\cite{Knox:2019rjx}. For the best-fit model in the middle panel Fig.~\ref{fig:fede}, over $50\%$ of the change in $r_{\star}$ occurs in the range $z=10^3$ to $z=10^4$, despite a much lower $f_\text{DE}(z)$. 

Beyond changes to the background, the dominant effect on the CMB power spectrum is radiation-driving due to the decay of the Weyl potential, similar to ADE~\cite{Lin:2019qug}. For a given $\Omega_i$, with all other components zero, there is an enhancement in the Weyl potential close to $a_i$, then a subsequent decay (leading to increased radiation-driving) as the dark energy perturbations stabilise due to $c_\mathrm{s}^2$. The ratio of the Weyl potential at $z_{\star}$, compared to a model with $\Omega_i=0$, tends to a constant for $k \gtrsim 0.1\, {\rm Mpc}^{-1}$. A similar effect is seen in the axion model and ADE. In the full spike model, however, an {\em increasing} $f_\text{DE}(z)$ at higher $z$ means that smaller scale modes ($k \gtrsim 0.1\, {\rm Mpc}^{-1}$) undergo more of a driving effect, resulting in increased power in the high $\ell$ CMB power spectrum. 

Although there is a large improvement in $\Delta \chi^2$, the p-value is only 0.85, due to the extra degrees of freedom we have introduced. %One simple method to quantify this is the Akaike information criterion (AIC)~\cite{1974ITAC...19..716A}, defined as ${\rm AIC}=\chi^2 + 2k$, where  $k$ is the number of parameters of the model. The AIC therefore disfavours the spike model over \lcdm. However, although useful as a guide, it is better to quantify model selection in terms of the Bayesian Evidence which we do in the next section. 
In the following section we introduce a correlation prior to reduce any spurious degrees of freedom and to calculate the Bayesian Evidence.

{\em Correlation Prior.---}%
 Fitting 32 spike components can cause slow convergence of MCMC, due to correlated, flat directions in the likelihood. Following~\citep{Zhao:2012aw}, we impose an explicit Gaussian prior on $\Delta_i$, which is unbounded.  This prior is specified by an amplitude and correlation length -- a larger correlation length favours smoother reconstructions, constraining flat, correlated directions. 
 
This approach has close similarities to Gaussian process reconstruction, and as such we use the Matern 5/2 covariance function, $C_{ij}^{\rm prior} = \sigma^2 \left( 1 + \frac{d_{ij}}{\rho} +  \frac{d_{ij}^2}{\rho^2} \right) \exp \left(-\frac{d_{ij}}{\rho} \right)$
where $d_{ij}=|\log_{10}a_i / a_j|$ and $\sigma$, and $\rho$ are hyper-parameters describing the amplitude and correlation length. A Gaussian process from the Matern 5/2 covariance is twice differentiable, so encourages continuous reconstructions of $w(z)$ and its derivative. As discussed in~\cite{Crittenden:2011aa}, the precise form of the covariance function is not too important, slightly shifting the spectrum of eigenmodes.

 The prior adds an additional term to the $\chi^2$ of a model, $\chi^2=\chi^2_{\rm data} + \chi^2_{\rm prior}$, where
$\chi^2_\mathrm{prior}=-2\ln {\cal P}_{\rm prior} = \mathrm{\Delta^T} ({\bf C}^{\rm prior})^{-1} \mathrm{\Delta}$. We numerically marginalise over the mean value of $\mathrm{\Delta}$ during the MCMC. This approach allows a straightforward estimation of the Bayesian evidence, $E \equiv \int d^n {\bf P} \, {\cal P}({\bf D| P})\, {\cal P}_{\rm prior}({\bf P})$. Assuming a Gaussian posterior covariance matrix, ${\mathbf C}^{\rm post}$, the evidence is~\citep{Zhao:2012aw}
\begin{equation}
E \propto \left(\frac{\det{\mathbf C}^{\rm post}}{\det{\mathbf C}^{\rm prior}}\right)^{1/2} e^{-\chi^2/2}\,.
\end{equation}
The first term is the fraction of prior parameter volume used by the data, and the second is the goodness-of-fit.

In order to choose the hyper-parameters $\sigma$ and $\rho$, we ran a coarse grid of models on mock data, assuming a fiducial axion $f_\text{DE}(z)$. A good compromise between minimal reconstruction bias and prior variance was found to be $\sigma=1.0$ and $\rho=0.75$. These hyper-parameters are then fixed in our MCMC analysis. 

Performing additional runs with this covariance prior, we now find a $\Delta \chi^2=-17.9$ improvement over  \lcdm\ for the baseline+ext dataset, and $\Delta \chi^2=-14.0$ with the inclusion of the DES prior. Parameters shift towards their preferred axion values, including $c_\mathrm{s}^2$. The reconstruction  of $f_\text{DE}(z)$ in Fig.~\ref{fig:fede} is now much closer to the axion, with an almost identical peak height and position. The covariance prior has heavily penalised the high $z$ peak, even though it can provide a better fit to CMB data. 

Overall, the difference in log-evidence of the spike model with a covariance and DES prior versus \lcdm\ is $\Delta \ln E=+5.0$, representing moderate evidence on the Jefferys' scale. Although this could potentially be tuned by changing $\sigma$ and $\rho$ (to avoid over penalising the high $z$ peak while maximising the available prior volume used), this is post-hoc, and it would be preferable to instead have a physical model.

{\em Conclusions.---}%
Resolving the Hubble tension is one of the most pressing issues in modern cosmology. In this letter, we have taken seriously the possibility that dark energy is dynamical \citep{Copeland:2006wr} in nature and rather than consider a particular model, we have performed a model-independent reconstruction from a redshift of $10^5$ to today. In that way we can ask whether it allows for the presence of EDE, whilst also being consistent with today's constraints on late time DE. The fascinating answer is that it can. With minimal prior assumptions, we find an improvement of $\Delta \chi^2=-41.3$ and $\Delta \chi^2=-25.1$ over \lcdm\ and the axion fluid respectively, for the baseline+ext dataset. Even including covariance priors on the model we still find moderate evidence that the Universe had a non-negligible amount of dark energy at a redshift $z \sim 10^5$, as well as around matter-radiation equality ($z \sim 3000$). The high $z$ component changes the radiation driving envelope that modifies the high $\ell$ CMB power spectrum, potentially alleviating the tension between \Planck\ and \textsc{ACT} data. 

We find the preferred sound speed is $c_s^2 \sim 1/3$. If this were the case, then the fluid could not be due to a standard quintessence type field evolution, which always gives $c_s^2 \sim 1$. It could in principle be some kind of additional radiation type matter, or there are a number of non-standard models which satisfy the sound speed constraint such as K-essence with a kinetic term proportional to $X^2$, where $X=-\dot{\phi}^2/2$, or models which directly couple the dark matter and dark energy (see \citep{Skordis:2015yra} for details). Of course, when we are dealing with evidence that is moderate, it means we have to be careful how much emphasis we place on the results. We accept that of course, but finish by noting that if there were new physics responsible for this occurring at a redshift of around $10^{4} - 10^5$, and with $c_s^2 \sim 1/3$, that would be a major result and worth exploring further.

{\em Acknowledgments.---}%
We thank Carsten van der Bruck, Simon Dye, Colin Hill, Lloyd Knox, Maggie Lieu, Marius Millea, Florian Niedermann, Levon Pogosian, Vivian Poulin for useful comments and discussions. TC is supported by an STFC studentship. EJC acknowledges support from STFC consolidated grant ST/T000732/1. AJM is supported by a Royal Society University Research Fellowship.

%\nocite{*}

\bibliographystyle{apsrev4-1}
\bibliography{bib}% Produces the bibliography via BibTeX.

\end{document}